\documentclass[%
 reprint,
 amsmath,amssymb,
 aps,
pra 
]{revtex4-2}
\usepackage{graphicx}%
\usepackage{color}
\usepackage{dcolumn}%
\usepackage{bm}%
\usepackage{booktabs}
\usepackage{hyperref}%
\usepackage{multirow}
\usepackage{makecell}

\begin{document}

\newcommand{\nn}{$\rm N_2$}
\newcommand{\hh}{$\rm H_2$}
\newcommand{\deiso}{$D_{\rm{e}, \lambda = 0}$}
\newcommand{\reiso}{$R_{\rm{e}, \lambda = 2}$}
\newcommand{\wvnum}{cm$^{-1}$}
\newcommand{\deo}{$D_{\text{e},\lambda=0}$}
\newcommand{\reo}{$R_{\text{e},\lambda=0}$}
\newcommand{\deaniso}{$D_{\rm{e}, \lambda = 2}$}
\newcommand{\reaniso}{$R_{\rm{e}, \lambda = 2}$}
\newcommand{\rbh}{$\text{Rb--H}_2$}
\newcommand{\rbn}{$\text{Rb--N}_2$}
\newcommand{\sigtot}{$\sigma_{\rm{tot}}$}
\newcommand{\colrate}[1]{$\kappa_{#1}$}
\newcommand{\rateunit}{$\times~10^{-15}~\rm{m}^3 / \rm{s}$}

\title{Invariance of quantum scattering rate coefficients to anisotropy of atom-molecule interactions }

\preprint{APS/123-QED}

\author{Xuyang Guo}
\email[Corresponding author: ]{xyguo@chem.ubc.ca}
 \affiliation{%
 Department of Chemistry, University of British Columbia,
Vancouver, B.C. V6T 1Z1, Canada%
 }%

 \author{Kirk W. Madison}%
 \affiliation{%
 Department of Physics and Astronomy, University of British Columbia, Vancouver, B.C. V6T 1Z1, Canada%
 }%

\author{James L. Booth}
\affiliation{
    Physics Department, British Columbia Institute of Technology,
   Burnaby, B.C. V5G 3H2, Canada%
}%

\author{Roman V. Krems}

\affiliation{Department of Chemistry \& Stewart Blusson Quantum Matter Institute, University of British Columbia,
Vancouver, B.C. V6T 1Z1, Canada 
}%

\date{\today}

\begin{abstract}
Quantum scattering calculations for strongly interacting molecular systems are computationally demanding due to the large number of molecular states coupled by the anisotropy of atom - molecule interactions. We demonstrate that thermal rate coefficients for total (elastic + inelastic) atom - molecule scattering are insensitive to the interaction anisotropy of the underlying potential energy surface. In particular, we show that the rate coefficients for \rbh~and \rbn~scattering at room temperature can be computed to 1\%  accuracy with anisotropy set to zero, reducing the complexity of coupled channel quantum scattering calculations to numerical solutions of a single differential equation. Our numerical calculations 
and statistical analysis based on Gaussian process regression elucidate the origin and limitations of the invariance of the total scattering rate coefficients to changes in atom - molecule interaction anisotropy.

\end{abstract}
\maketitle 

\section{Introduction} 
\label{sec:intro}
Quantum scattering calculations of rate coefficients for molecular collisions are widely used as benchmarks of experimental measurements~\cite{son2020collisional,Wiesenfeld2021,liu2021precision,Hong2021,paliwal2021determining}, to elucidate mechanisms of energy transfer observed in molecular dynamics experiments~\cite{paliwal2021determining,geistfeld2020qct,besemer2022glory,mukherjee2023quantum,mandal2023adiabatic}, and for astrophysical models~\cite{Faure2022,Boursier2020,Her2021,Mandal2023,Gon2024}. 
Total (elastic + inelastic) scattering rate coefficients are essential inputs into methods using laser-cooled atoms to determine ambient gas density under high and ultra-high vacuum conditions via collision-induced heating and trap loss measurements~\cite{madison2012cold,booth54METHODDEVICE2014,boothUniversalityQuantumDiffractive2019,shenRealizationUniversalQuantum2020,shenRefiningColdAtom2021,barkerAccurateMeasurementLoss2023,shenCrosscalibrationAtomicPressure2023,stewartMeasurementRbRbVan2022,frielingCrosscalibrationQuantumAtomic2024,Avinash2024}. For this sensor application, it is desired that the uncertainty of the rate coefficients is below 1\%~\cite{boothUniversalityQuantumDiffractive2019}. 
It is also desirable for these experiments to know the uncertainty of the computed collision rate coefficients. 
The uncertainty of the rate coefficients can be estimated by repeated quantum scattering calculations with underlying potential energy surfaces (PES) varied within their errors. 
However, quantum scattering calculations for strongly interacting molecular systems are computationally demanding due to the large number of molecular states coupled by the anisotropy of atom - molecule interactions. For example, rigorous quantum scattering calculations of room-temperature rate coefficients for Li--\nn~and \rbn~collisions relevant for vacuum metrology require as many as 742 rotational states coupled by the interaction anisotropy for convergence~\cite{klosElasticGlancingangleRate2023}.

Recent work has investigated the probability distributions of cross sections for atom - molecule scattering that result from a distribution of interaction potential inputs to the calculations~\cite{Cuijie2013Elastic,moritaUniversalProbabilityDistributions2019}. Ref.~\cite{moritaUniversalProbabilityDistributions2019} showed that reliable probabilistic predictions can be obtained with significantly reduced numerical effort.  This result is relevant for the task of determining the uncertainty of a computed collision rate coefficient and demonstrates the feasibility of making predictions of experimentally relevant observables for certain systems previously considered out of reach of quantum dynamics theory.  In other work motivated by vacuum metrology, it was found that certain, thermally averaged, atomic and molecular collision observables are invariant, or weakly sensitive, to changes of underlying interaction potentials at short inter-particle distances~\cite{boothUniversalityQuantumDiffractive2019,shenRealizationUniversalQuantum2020,shenRefiningColdAtom2021,boothRevisingUniversalityHypothesis2024}. Ref.~\cite{guo2024boundariesuniversalitythermalcollisions} identified the range of parameters over which atom-atom collision rate coefficients exhibit this type of short-range invariance.  Together, these results can be exploited to reduce the complexity of quantum scattering calculations without compromising their accuracy.

In the present work, we show that thermal rate coefficients for total atom - molecule scattering are invariant or weakly sensitive to changes in the interaction anisotropy. As a result, accurate rate coefficients for total scattering can be computed with single-channel quantum scattering calculations, using only the isotropic contribution of the atom - molecule interaction potential. We demonstrate that this invariance stems from thermal averaging, consistent with the results for quantum diffractive scattering in Refs.~\cite{boothUniversalityQuantumDiffractive2019,boothRevisingUniversalityHypothesis2024,guo2024boundariesuniversalitythermalcollisions}. 
We perform a comprehensive analysis to examine the response of total thermal collision rate coefficients to the variations in both short-range and long-range parts of anisotropic potentials.
This work is based on rigorous quantum scattering calculations employing the most accurate available interaction potentials for \rbn\ and \rbh\ collisions. 
We illustrate the invariance by treating the collision observables as probabilistic predictions determined by a distribution of interaction potentials~\cite{moritaUniversalProbabilityDistributions2019,guo2024boundariesuniversalitythermalcollisions}. 
We show that the distributions of total rate coefficients are largely insensitive to changes of anisotropy, yielding similar collision rate coefficients over a wide range of interaction potentials for \rbh\ and \rbn\ thermal collisions. 
We explore the boundaries of the invariance with respect to the strength of anisotropy of the interaction potentials. The observed boundaries significantly exceed typical uncertainties in anisotropic potentials from electronic structure calculations~\cite{dereviankoElectricDipolePolarizabilities2010,klosElasticGlancingangleRate2023}. 
Finally, we examine in detail the difference between multi-channel and single-channel calculations of total collision rate coefficients.
We show that the difference is smaller than 1\% of the rate coefficient magnitudes even for the atom-molecule collision systems with strong anisotropy and inelastic scattering.

\section{Method}
\label{sec:meth}
\subsection{Scattering calculations}
\label{sec:cc}
We calculate and analyze rate coefficients for collisions between trapped $^{87}$Rb atoms and homonuclear molecules ($^2$\hh\ and $^{28}$\nn) in a thermal background gas at temperature $T$. We treat the diatomic molecules as rigid rotors initially in a rotational state with a fixed value $j$ of the rotational angular momentum. 
The rate coefficients $\kappa_j(T)$ are computed by integrating the cross sections $\sigma_j(\tilde{E})$  with the Maxwell-Boltzmann distribution
\begin{equation}
    \kappa_j(T)=  \sqrt{\frac{8 k_{\rm B} T}{\pi m}} \frac{1}{\left(k_{\rm B} T\right)^2}
    \int_0^{\infty}  E \sigma_j(\tilde{E}) \exp{\left(-\frac{E}{k_{\rm B}T}\right)} \mathrm{d}E,
    \label{eq:cal_rate}
\end{equation} where $k_{\rm B}$ is the Boltzmann constant, $m$ is the mass of the molecule, $E$ is the kinetic energy of the incident particle, $\tilde{E}=\mu E/m$ is the collision energy, and $\mu$ is the reduced mass of the collision complex. 
Note that we use $\tilde E$ as the argument of the cross sections and the mass $m$ of the molecules in the background gas for thermal averaging. 
This is appropriate for collisions of molecules from a thermal gas at temperature $T$ with trapped atoms assumed to be stationary. 
The kinetic energy of the trapped atoms considered here is $\sim 10^{6}$ times smaller than that of the background gas molecules and can, therefore, be ignored~\cite{Avinash2024}.

It is instructive to make an explicit connection between thermal averaging given by Eq.~(\ref{eq:cal_rate}) and the cumulative probability distribution (CPD) approach used in Ref.~\cite{moritaUniversalProbabilityDistributions2019}. 
The CPD approach in Ref.~\cite{moritaUniversalProbabilityDistributions2019} constructs a cumulative probability distribution for an observable derived from the energy-dependent scattering data by treating the scaling parameter of the interaction potential as a random variable and integrating over it. The rate coefficient defined by Eq.~\eqref{eq:cal_rate} is a functional of the energy-dependent cross section $\sigma(E)$. We write $\kappa_j(T;\theta)$ to emphasize this dependence for a given potential parameter vector $\theta$. The CPD of the rate is then the probability
\begin{eqnarray}
\mathrm{CPD}_{\kappa_j}(\kappa_{j,0})&&=\Pr_\theta\!\big( \kappa_j(T;\theta) \le \kappa_{j,0} \big) \\ \nonumber
          &&=\; \int q\big( \kappa_j(T;\theta)\le \kappa_{j,0}\big)\, p(\theta)\, d\theta,
\end{eqnarray}
where $p(\theta)$ is the (prior) distribution over potential parameters and $q$ is the indicator function. Equivalently, the CPD for the cross section at fixed $E$ is $\mathrm{CPD}_\sigma(s;E)=\Pr_\theta\!\big(\sigma(E;\theta)\le s\big)$. The CPD for $\kappa_j(T)$ thus follows from integrating the CPDs of $\sigma(E;\theta)$ with the Maxwell-Boltzmann weights given by $E\,e^{-E/(k_B T)}$~\cite{1992hermann} in Eq.~\eqref{eq:cal_rate}. The CPD definition in Ref.~\cite{moritaUniversalProbabilityDistributions2019} naturally maps onto the present integral of thermal rate. Thermal averaging is a linear functional of $\sigma(E)$ and marginalizing over $\theta$ produces the CPD for $\kappa_j(T)$.

We use rigorous close-coupling (CC) calculations to compute $\sigma_j(\tilde{E})$.  The CC approach is well documented elsewhere~\cite{arthursTheoryScatteringRigid1960,lester1971calculation,alexander1991quantum,manolopoulos1992quantum,bernstein2013atom}. Here, we present a brief description of the most relevant details. 
The collision cross sections $\sigma_j(\tilde{E})$ are obtained from the numerical solutions of the time-independent Schrödinger equation represented as a system of coupled differential equations
\begin{eqnarray}
    \left[\frac{\mathrm{d}^2}{\mathrm{~d}R^2}-\frac{l^{\prime}\left(l^{\prime}+1\right)}{R^2}+k_{j^{\prime}}^2\right] F_{j^{\prime} l^{\prime}}^{Jjl}(R) = \nonumber \\
    \sum_{j^{\prime \prime}} \sum_{l^{\prime \prime}}U_{j^{\prime \prime} l^{\prime \prime} ; j^{\prime} l^{\prime}}^J F_{j^{\prime \prime} l^{\prime \prime}}^{Jjl} (R),
\label{cc}
\end{eqnarray}
where $R$ is the atom-molecule distance, $l$ is the orbital angular momentum of the collision complex, $j$ is the rotational angular momentum of the molecule, $J=j+l$ is the total angular momentum, and the matrix elements $U_{j^{\prime \prime} l^{\prime \prime}; j^{\prime} l^{\prime}}^J$ are determined by the atom - molecule interaction potential described in Section~\ref{sec:pes}. 
The nuclear-spin statistics restrict allowed rotational levels for homonuclear diatoms. For the ground electronic states X$^1\Sigma_g^+$ of both N$_2$ and H$_2$, the nuclear-spin symmetry imposes restrictions on the allowed rotational levels. 
The para-H$_2$ (total nuclear spin $I=0$) is restricted to even $j$, while the ortho-H$_2$ ($I=1$) corresponds to odd $j$. 
The ortho-N$_2$ (total nuclear spin $I=0$ or $2$) is restricted to even $j$, whereas the para-N$_2$ ($I=1$) corresponds to odd $j$. 
The scattering results for H$_2$ and N$_2$ reported here are for the vibrational ground state $v=0$. 
The wave number $k_{j^\prime}$ is defined as
\begin{eqnarray}
    k_{ j^\prime}^2=\frac{2 \mu}{\hbar^2}\left[E_{\rm total}- B_{\rm r} j^\prime\left(j^\prime+1\right)\right],
\end{eqnarray} where $B_{\rm r}$ is the rotational constant  of the diatomic molecule, and $E_{\rm total}$ is the total energy of the collision complex.

We use the log-derivative method~\cite{manolopoulosImprovedLogDerivative1986} to integrate the coupled equations~(\ref{cc}) and match the numerical solutions to the scattering boundary conditions, yielding the scattering matrix $S$, as described, for example, in Ref.~\cite{JOHNSON1973445,MANOLOPOULOS198823}. The state-resolved cross sections for the $j \rightarrow j^\prime$ transitions in atom - molecule collisions are then computed as
\begin{eqnarray}
\sigma_{j j^{\prime}}(\tilde{E})=&&\frac{4\pi}{(2 j+1) k_{j}^2} \sum_{J=0}^{\infty} \sum_{l=|J-j| }^{J+j} \sum_{l^{\prime}=\left|J-j^{\prime}\right|}^{J+j^{\prime}}(2 J+1) \nonumber \\
&&\times \left|\delta_{j j^{\prime}} \delta_{ll^{\prime}}-S^J\left(j l ; j^{\prime} l^{\prime}\right)\right|^2,
\label{partial-wave}
\end{eqnarray} where $j=j^{\prime}$ and $j \ne j^{\prime}$ corresponds to elastic and inelastic scattering, respectively.
The inelastic cross sections in Eq. (\ref{eq:cal_rate}) are computed for an initial rotational state $j$ as
\begin{eqnarray}
\label{inelastic-cross-section}
\sigma_j(\tilde{E})  = \sum_{j^\prime} \sigma_{jj^{\prime}}(\tilde{E}), 
\end{eqnarray} where the sum is over all $ j^{\prime} \neq j$ states allowed by total energy. The sum in Eq.~(\ref{inelastic-cross-section}) thus includes both excitation and de-excitation transitions.

We allow the maximum rotational angular momentum $j_{\text{max}}$ to be 6 for $^2 \rm {H}_2$ and 26 for $^{28}\rm{N}_2$. This ensures convergence of cross sections at the highest collision energy considered to within $\leq 0.1\%$. 
We include enough terms in the partial wave sum (\ref{partial-wave}) to ensure convergence of cross sections to $\leq 0.1\%$ and extend the range of incident particle energy in Eq.~(\ref{eq:cal_rate}) to ensure convergence of the thermal rate coefficients to  $\leq 0.25\%$. The number of partial waves required for convergence depends on the reduced mass of the collision complex and the collision energy $\tilde{E}$. We include quantum states with $l \leq 200$ for \rbh\ collisions and $\leq 650$ for \rbn\ collisions.

For further analysis, we also calculate the rate coefficients $\kappa_j(T)$ assuming the partial cross sections are determined solely by elastic scattering from the purely isotropic, long-range part of the interaction potential. The corresponding scattering phase shifts obtained from the semi-classical Jeffreys-Born approximation~\cite{boothRevisingUniversalityHypothesis2024}:
\begin{eqnarray}
    \eta_l(k) = && \frac{3\pi}{16}\left(\frac{\mu C_6}{\hbar^2}\right)\frac{k^4}{(l+\frac{1}{2})^5} + \frac{5\pi}{32}\left(\frac{\mu C_8}{\hbar^2}\right)\frac{k^6}{(l+\frac{1}{2})^7} \nonumber\\
    && + \frac{35\pi}{256}\left(\frac{\mu C_{10}}{\hbar^2}\right)\frac{k^8}{(l+\frac{1}{2})^9},
    \label{eq:JBJB}
\end{eqnarray} where $C_n$ are the coefficients in the isotropic, long-range expansion of the interaction potential:
\begin{eqnarray}
\lim_{R \rightarrow \infty} V(R) = -\frac{C_6}{R^6} - \frac{C_8}{R^8} - \frac{C_{10}}{R^{10}}.
\label{LR}
\end{eqnarray} The cross-section in Eq.~\ref{partial-wave} can then be simplified to:
\begin{equation}
    \sigma(\tilde{E}) = \frac{4\pi}{k^2} \sum_{l=0}^{\infty} (2l+1) \sin^2 \eta_l.
    \label{partial-wave-single}
\end{equation}
The resulting rate coefficients are determined solely by the elastic long-range part of interaction potentials.

\subsection{Atom-molecule interaction potential}

\label{sec:pes}

Within the rigid rotor approximation for the diatomic molecule, the atom - molecule interaction potential is a function of two coordinates. We use the Jacobi coordinates $(R, \theta)$ defined as 
the magnitude $R$ of vector $\bm R$ joining the centers of mass of the atom and the diatomic molecule and the angle $\theta$ 
 between the axis of the diatomic molecule and $\bm R$. 
The atom-molecule interaction potential  $V(R, \theta)$ is represented as a sum of Legendre polynomials
\begin{equation}
    V(R, \theta) = \sum_{\lambda=0}^{\infty} V_\lambda(R) P_\lambda(\cos \theta).
 \label{legendre}
\end{equation}
For homonuclear diatomic molecules, the expansion contains only terms with even values of $\lambda$.
The expansion coefficients $V_{\lambda}(R)$ are computed using the global PES from Ref. 
\cite{upadhyayEnhancedSpinCoherence2019} for \rbh\  and Ref. \cite{klosElasticGlancingangleRate2023} for \rbn . 
We include terms with $\lambda=0,2,4,$ and $6$ in Eq. ({\ref{legendre}}) for both interaction potentials considered here. 
The leading expansion coefficient $V_0$ characterizes the isotropic part of the interaction potential, which predominantly determines elastic scattering. The expansion coefficients $V_{\lambda}$  with $\lambda > 0$ describe the anisotropy of the atom - molecule interaction that leads to inelastic scattering.

This work aims to investigate the response of the thermal collision rate coefficients in Eq.~(\ref{eq:cal_rate}) to the variations in the isotropic and anisotropic parts of the interaction potentials at short range $R$. To accomplish this, we model $V_\lambda(R)$ using Morse long-range (MLR) functions~\cite{medvedevInitioInteratomicPotentials2018}
        \begin{equation}
        V_{\lambda}(R)=D_{\mathrm{e}, \lambda}\mathcal{F}_\lambda(R),
        \label{eq:mlr}
    \end{equation} 
    where 
\begin{equation}
    \mathcal{F}_\lambda(R) = \left(1-\frac{u_{\mathrm{LR}}(R)}{u_{\mathrm{LR}}\left(R_{\mathrm{e}, \lambda}\right)} \exp \left[-\beta(R) \cdot y_p^{\mathrm{eq}}(R)\right]\right)^2,
    \label{fterm}
\end{equation}       
    $D_{\mathrm{e}, \lambda}$ is the well depth, $R_{{\rm e}, \lambda}$ is the equilibrium distance of the corresponding expansion coefficient. 
    The function
    $\beta(R)$ is defined as a polynomial function
    \begin{equation}
        \beta(R)=y_p^{\mathrm{ref}} \beta_{\infty}+\left[1-y_p^{\mathrm{ref}}\right] \sum_{i=0}^N \beta_i\left[y_q^{\mathrm{ref}}\right]^i
        \label{eq:ref_dist}
        \end{equation} of the radial variables, or reduced coordinates $y_{p, q}^{\mathrm{ref}}(R)$
    \begin{equation}
    y_{p, q}^{\mathrm{ref}}(R)=\frac{R^{p, q}-R_{\mathrm{ref}}^{p, q}}{R^{p, q}+R_{\mathrm{ref}}^{p, q}}
    \end{equation} where $R_{\mathrm{ref}}$ is the reference distance \cite{leroyAccurateAnalyticPotentials2009}. The integer powers $p$ and $q$ are fixed to 4 and 5, respectively, following Ref.~\cite{medvedevInitioInteratomicPotentials2018}. The radial variable $y_p^{\mathrm{eq}}$ takes the form of Eq.~(\ref{eq:ref_dist}) with $R_{\mathrm{ref}}=R_e$. The constraint parameter $\beta_{\infty}$ is fixed to $\ln \left\{2 D_{\mathrm{e}} / u_{\mathrm{LR}}\left(R_{\mathrm{e}}\right)\right\}$. The long-range potential $u_{\mathrm{LR}}$ in Eq.~(\ref{eq:mlr}) has the asymptotic form
    \begin{equation}
        u_{\mathrm{LR}}(R)= D_6 \frac{C_6}{R^6} + D_8 \frac{C_8}{R^8}+ D_{10} \frac{C_{10}}{R^{10}},
        \end{equation} where $D_n$ is the damping function defined as the generalized Douketis function \cite{douketisIntermolecularForcesHybrid1982a,leroyLongrangeDampingFunctions2011}
        \begin{equation}
            D_n(r,s)=\left(1-\exp \left(-\frac{b(s)(\rho R)}{n}-\frac{c(s)(\rho R)^2}{n^{1 / 2}}\right)\right)^{n+s},
            \end{equation} where $b(s)$ and $c(s)$ are system-independent parameters determined by the choice of $s$. In this work, we choose $s=-1$, $b = 3.3$, and $c=0.423$, following Ref.~\cite{medvedevInitioInteratomicPotentials2018}. The dimensionless parameter \(\rho = \frac{2 \rho_{\mathrm{M}} \rho_{\mathrm{BG}}}{\left( \rho_{\mathrm{M}} + \rho_{\mathrm{BG}} \right)}\) \cite{douketisIntermolecularForcesHybrid1982a} is calculated as the harmonic mean of the atomic counterparts \(\rho_i = \left( \frac{\mathrm{IP}_{\mathrm{i}}}{\mathrm{IP}_{\mathrm{H}}} \right)^{\frac{2}{3}}\), where \(\mathrm{IP}_{\mathrm{i}}\) and \(\mathrm{IP}_{\mathrm{H}}\) represent the ionization potentials of the respective molecule and atomic hydrogen. We fit the expansion coefficients of the \textit{ab initio} potentials in Ref.~\cite{klosElasticGlancingangleRate2023} and \cite{upadhyayEnhancedSpinCoherence2019} by the non-linear least squares method to obtain the MLR parameters. Compared to the \textit{ab intio} energies, the root mean squared error of the resulting fits is smaller than $1~\mathrm{cm}^{-1}$.

\subsection{Statistical analysis}
\label{sec:stat}
\begin{figure}[b]
    \includegraphics[width=0.5\textwidth]{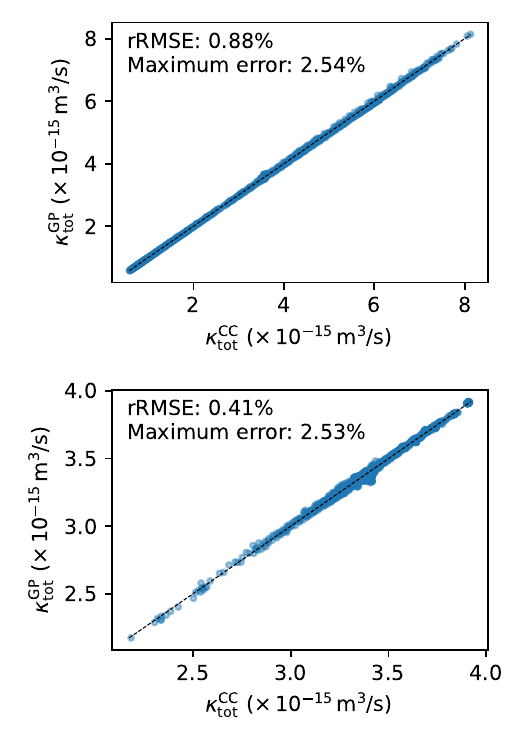}
\caption{Comparison of $\kappa_{\text{tot}}$ predicted by the 5D GP model with the results of CC calculations. The CC calculations for training and testing the GP models are performed with the reduced mass, $C_{6,\lambda=0}$ and $R_{\text{e}, \lambda = 0}$ corresponding to the \rbh\ (upper) and \rbn\ (lower) collision systems. The rRMSE (\%) in Eq.~\ref{rRMSE} and maximum error (\%) of the GP predictions relative to the results of CC calculations are indicated in each panel.}
\label{figure1}
\end{figure}

In addition to rigorous CC calculations, we adopt a statistical analysis approach to examine the response of total collision rate coefficients to variations in both the short-range part and long-range part of the interaction anisotropy as well as changes in the magnitude of the rotational constant of the diatomic molecule. 
We begin by treating the equilibrium distance \(R_{\text{e}, \lambda = 2}\), the well depth magnitude \(D_{\text{e}, \lambda = 2}\) and the long-range interaction coefficient $C_{6,\lambda=2}$ as variables with wide ranges of values, leading to a distribution of interaction potentials of varying anisotropy. 
To ensure the generality of conclusions, we also vary $D_{\text{e},\lambda=0}$ and the rotational constant $B_{\rm r}$ of the molecule. 
The rate coefficients $\kappa_{\rm tot}$ are thus considered to be functions of five variables.

To obtain the global dependence of $\kappa_{\rm tot}$ on these five variables, we build a Gaussian process (GP) regression model of $\kappa_{\rm tot}$  trained by rigorous multi-channel scattering calculations at randomly sampled combinations of $R_{\text{e}, \lambda = 2}, D_{\text{e}, \lambda = 2}, C_{6,\lambda=2}, D_{\text{e},\lambda=0}$, and $B_{\rm r}$. 
GP regression is a supervised learning algorithm that uses a set of $N$ input-output pairs $\mathcal{D} = \{ (\bm{x}_i, {y}_i) \}_{i=1}^{N}$ to infer a target function $y(\bm x)$. 
In the present work, $\kappa_{\rm tot} \Rightarrow y$ and $( R_{\text{e}, \lambda = 2}, D_{\text{e}, \lambda = 2}, C_{6,\lambda=2}, D_{\text{e},\lambda=0}, B_{\rm r} ) \Rightarrow \bm x \in \mathbb{R}^5$. 
Once trained, a GP model can be used to predict the value of $y$ at an arbitrary point $\bm{x}_*$ of input space as follows\cite{rasmussen2005gaussian}: 
\begin{equation}
	{{y}}(\bm{x}_*) = \bm{k}^\top(\bm{x}_*) (\mathbf{K} + \sigma_n^2 \bm{I})^{-1} \bm{y}, 
\label{eq:gp_pred}
\end{equation} where $\bm y$ is a vector collecting $N$ training points $y_i$, 
\(\bm{k}\) is the vector of size $N$ with elements given by the values of a kernel function $k(\bm x_i, \bm x_*)$ and \(\mathbf{K}\) is the $N \times N$ matrix with elements given by the values of the kernel function $k(\bm x_i, \bm x_j)$ for $ \bm x_i, \bm x_j \in {\cal D}$. 
The value of \(\sigma_n^2\) is set to zero because the training points $y_i$ are the results of the CC calculations, which produce noiseless data.  
The kernel function is chosen to be 
\begin{align}
    k(\bm{x}_i, \bm{x}_j) &= 
    A_1^2 \left(1 + \frac{\sqrt{3} r}{\ell_1} \right) \exp\left( -\frac{\sqrt{3} r}{\ell_1} \right) \nonumber \\ &\quad  +
    A_2^2 \left(1 + \frac{\sqrt{5} r}{\ell_2} + \frac{5 r^2}{3 \ell_2^2} \right) \exp\left( -\frac{\sqrt{5} r}{\ell_2} \right) \nonumber \\
    &\quad + A_3^2 \exp\left( -\frac{r^2}{2 \ell_3^2} \right) +
    A_4^2 \left(1 + \frac{r^2}{2 \alpha \ell_4^2} \right)^{-\alpha}\nonumber \\ &\quad +
    A_5^2 \left( \bm{x}_i^\top \bm{x}_i \right)
\end{align} where \( r = \|\bm{x}_i - \bm{x}_j\| \) denotes the Euclidean distance between the input data points, \( \ell_i \) are the characteristic length-scale parameters, \( \alpha \) is the shape parameter of the rational quadratic (RQ) kernel, and \( A_i \) are the amplitudes of different components of the kernel function.
The parameters of the kernel function are estimated by type-II maximum likelihood estimation \cite{rasmussen2005gaussian}.
The accuracy of the resulting GP models is illustrated in Fig. \ref{figure1}. Fig. \ref{figure1} includes rate coefficients from the entire range of the input parameters and shows the error of the GP model predictions as the departure from the diagonal broken line. 
We report the relative root-mean-square error (rRMSE) of the GP models, defined as \begin{eqnarray}
\operatorname{rRMSE}= \frac{\sqrt{\frac{1}{\mathcal{M}} \sum_i\left(\kappa_{\text{tot},i}^{\mathrm{GP}}-\kappa_{\text{tot},i}^{\mathrm{CC}}\right)^2}}{\frac{1}{\mathcal{M}} \sum_i \kappa_{\text{tot},i}^{\mathrm{CC}}} \times 100\%.
\label{rRMSE}
\end{eqnarray} where $\mathcal{M}$ is the number of training points.
The models illustrated in  Fig. \ref{figure1} are trained by 3300 (upper panel) and 1500 (lower panel) CC calculations of $\kappa_{\rm tot}$, and produce a maximum error of 2.5\% and a rRMSE of 0.9\% over randomly sampled test sets including 1900 (upper panel) and 3700 (lower panel) values of calculated $\kappa_{\rm tot}$.  We consider the following ranges of the input variables: 
$R_{\mathrm{e},\lambda=2} \in \left[ 0.25, 7 \right]~\text{\AA}, \quad D_{\mathrm{e},\lambda=2} \in \left[1,110\right]~\text{cm}^{-1}, \quad C_{6,\lambda=2} \in \left[0.5, 350\right]~E_h \cdot a_0^6, \quad D_{\mathrm{e},\lambda=0} \in \left[ 1, 110 \right]~\text{cm}^{-1}, \quad B_{\rm r} \in \left[0,10\right]~\text{cm}^{-1}.$

Given the GP model of the rate coefficients ${\kappa}_{\rm tot} = f(R_{\text{e}, \lambda = 2}, D_{\text{e}, \lambda = 2}, C_{6,\lambda=2}, D_{\text{e},\lambda=0}, B_{\rm r} )$, 
we treat a subset of the variables as random variables in order to examine the dependence of $\kappa_{\rm tot}$ 
on the remaining variables. We consider GP model predictions marginalized over the random variables as well as predictions for multiple, fixed samples of the random variables. Marginalization refers to integrating the predictive distribution from the GP model over the distribution of the random variables, thereby incorporating the variation in those variables into the analysis of the dependence of $\kappa_{\rm tot}$ on the remaining variables.
This allows for a comprehensive analysis of the dependence of $\kappa_{\rm tot}$ on the interaction anisotropy, accounting for the effects of variations in the strength of the isotropic part of the interaction potential and the rotational constant of the molecule on the resulting conclusions. The GP models are used to supplement direct multi-channel scattering calculations with specific subsets of the variables considered here. In these calculations, the remaining variables are kept fixed, which corresponds to examining cuts in the five-dimensional dependence of $\kappa_{\rm tot}$ on the interaction potential parameters and the rotational constant. 

\begin{table}[!htbp]
\caption{
\label{tab:rate}  
Rate coefficients (in units of $10^{-15}~\mathrm{m^3/s}$) for elastic ($\kappa_{\rm el}$), inelastic ($\kappa_{\rm inel}$) and total ($\kappa_{\rm tot}$) scattering of Rb with H$_2$ and N$_2$ initially in the rotational state $j$. The temperature is $T = 294.15$ K. 
The results for $\kappa_{\rm el}$, $\kappa_{\rm inel}$, and $\kappa_{\rm tot}$ are from cross sections computed with converged CC calculations and full PESs. 
The results for $\kappa_{\rm single}$ are from the single-channel differential equation with the corresponding isotropic part of the full PES obtained by setting  
all $V_{\lambda > 0}(R)$ in Eq. (\ref{legendre}) to zero. The results for $\kappa_{\rm SC}$ are computed using the semi-classical Jeffreys-Born approximation~\ref{eq:JBJB} with the long-range expansion coefficients for $V_{\lambda = 0}(R\rightarrow \infty)$. }
\begin{ruledtabular}

\begin{tabular}{lcccccc}
System & $j$ & $\kappa_\mathrm{el}$ & $\kappa_\mathrm{inel}$ & $\kappa_\mathrm{tot}$ & $\kappa_\mathrm{single}$ & $\kappa_\mathrm{SC}$  \\
\hline
\multirow{7}{*}{\rbh} 
 & 0 & 3.575 & 0.0003 & 3.576 & \multirow{7}{*}{3.567} & \multirow{7}{*}{5.964}\\
 & 1 & 3.578 & 0.0005 & 3.579 & &  \\
 & 2 & 3.577 & 0.0004 & 3.578 & &  \\
 & 3 & 3.577 & 0.0002 & 3.577 & &  \\
 & 4 & 3.577 & 0.0001 & 3.577 & &  \\
 & 5 & 3.577 & 0.00003 & 3.577 & &  \\
 & 6 & 3.577 & 0.00001 & 3.577 & &  \\
\hline
\multirow{7}{*}{\rbn} 
 & 0 & 3.107 & 0.306 & 3.413 & \multirow{7}{*}{3.395} & \multirow{7}{*}{3.467} \\
 & 1 & 2.937 & 0.488 & 3.425 & & \\
 & 2 & 3.287 & 0.109 & 3.396 & & \\
 & 3 & 3.302 & 0.120 & 3.422 & & \\
 & 4 & 3.341 & 0.058 & 3.399 & &  \\
 & 5 & 3.354 & 0.069 & 3.423 & &  \\
 & 6 & 3.370 & 0.057 & 3.427 & & \\
\end{tabular}
\end{ruledtabular}
\end{table}

\begin{table}[h]
\centering

\caption{\label{tab:popavg} Comparison of thermally averaged rate coefficients~(\ref{eq:popavg}) (in units of $10^{-15}~\mathrm{m^3/s}$) computed in the present work with results of experimental measurements~\cite{shenRealizationUniversalQuantum2020,barkerAccurateMeasurementLoss2023} or previous calculations~\cite{shenCrosscalibrationAtomicPressure2023,klosElasticGlancingangleRate2023}. }
\begin{tabular}{lcc}
\hline\hline
System & $\bar{\kappa}$ & $\bar{\kappa}_\mathrm{ref}$ \\
\hline
Rb-H$_2$ & 3.577
           & \makecell{ 3.574~\cite{shenCrosscalibrationAtomicPressure2023} \\ 3.89(10)~\cite{klosElasticGlancingangleRate2023}}\\
\hline
Rb-N$_2$ & 3.416 
           & \makecell{ 3.11(6)~\cite{shenRealizationUniversalQuantum2020}  \\ 3.6(1)~\cite{barkerAccurateMeasurementLoss2023} \\3.45(6)~\cite{klosElasticGlancingangleRate2023}}\\
\hline\hline
\end{tabular}
\end{table}

\begin{figure}[b]
    \centering
    \includegraphics[width=0.48\textwidth]{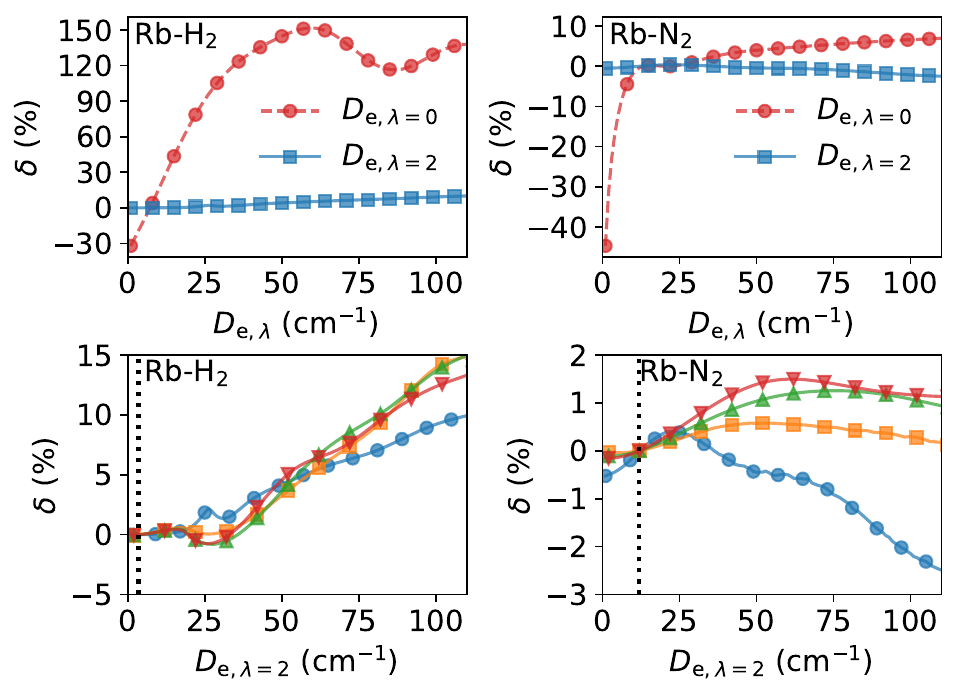}
\caption{CC calculations of the relative deviation (17) for collisions of Rb with H$_2$ (left) and N$_2$ (right). Upper panels: Relative deviation (\ref{eq:rel_err}) for collisions of molecules in the ground rotational state as functions of the well depth magnitude for the isotropic $V_{\lambda=0}$ (circles) and leading anisotropic $V_{\lambda=2}$ (squares) Legendre expansion coefficients of the full PES. Lower panels: The relative deviation~(\ref{eq:rel_err}) as functions of $D_{\mathrm{e},\lambda=2}$ for collisions of molecules in the rotational states $j=0$ (circles), $j=2$ (squares), $j=4$ (triangles), $j=6$ (inverted triangles). The reference values of $D_{\mathrm{e},\lambda=2}$ given by the MLR fits (\ref{eq:mlr}) are shown by the vertical dotted lines.}
\label{figure2}
\end{figure}
\begin{figure}[b]
    \includegraphics[width=0.48\textwidth]{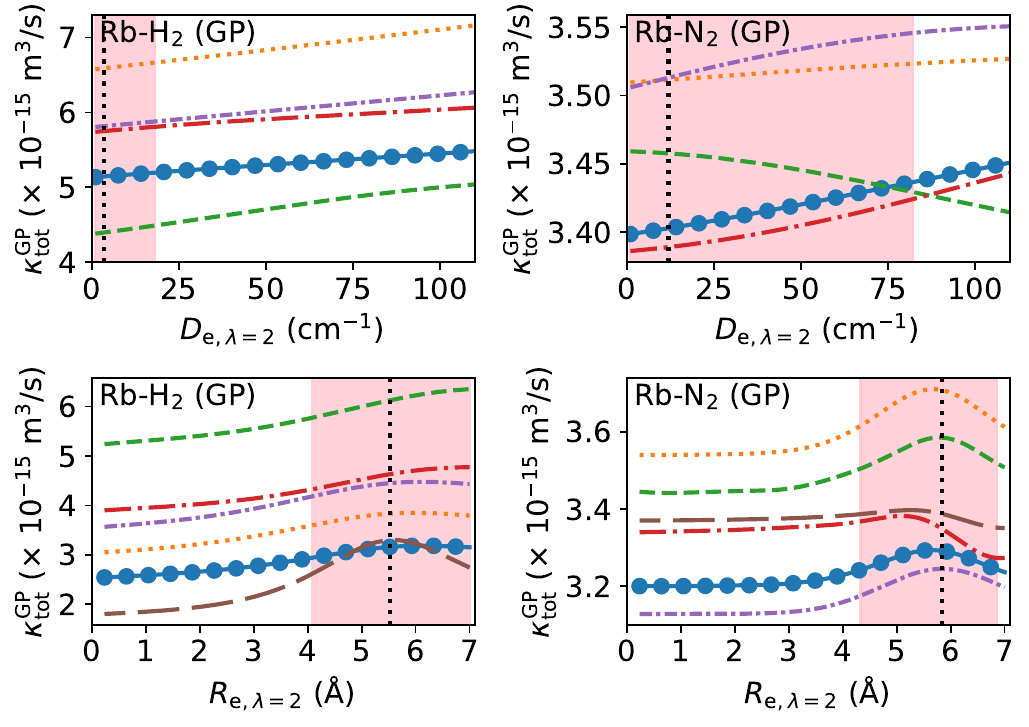}
\caption{
   Rate coefficients $\kappa_{\rm{tot}}$ from the GP models as functions of $D_{\mathrm{e},\lambda=2}$ (upper panels) and $R_{\text{e}, \lambda = 2}$ (lower panels), with the corresponding reference values from the MLR fits (\ref{eq:mlr}) shown by the vertical dotted lines. The change of $\kappa_{\rm{tot}}$ within 1\% is shown by the shaded regions. 
   The GP models are trained by the CC calculations performed with the reduced mass, $C_{6,\lambda=0}$ and $R_{{\rm e},\lambda=0}$ corresponding to the \rbh\ (left) and \rbn\ (right) collision systems.
  The lines with the symbols represent $\kappa_{\rm{tot}}$ marginalized over the variables not shown on the $x$-axis.
   The dashed and dotted lines without symbols show samples of ${\kappa}_{\rm tot} = f(R_{\text{e}, \lambda = 2}, D_{\text{e}, \lambda = 2}, C_{6,\lambda=2}, D_{\text{e},\lambda=0}, B_{\rm r} )$ for random combinations of the variables not shown on the $x$-axis.
}
\label{figure3}
\end{figure}

\begin{figure}[b]
    \includegraphics[width=0.4\textwidth]{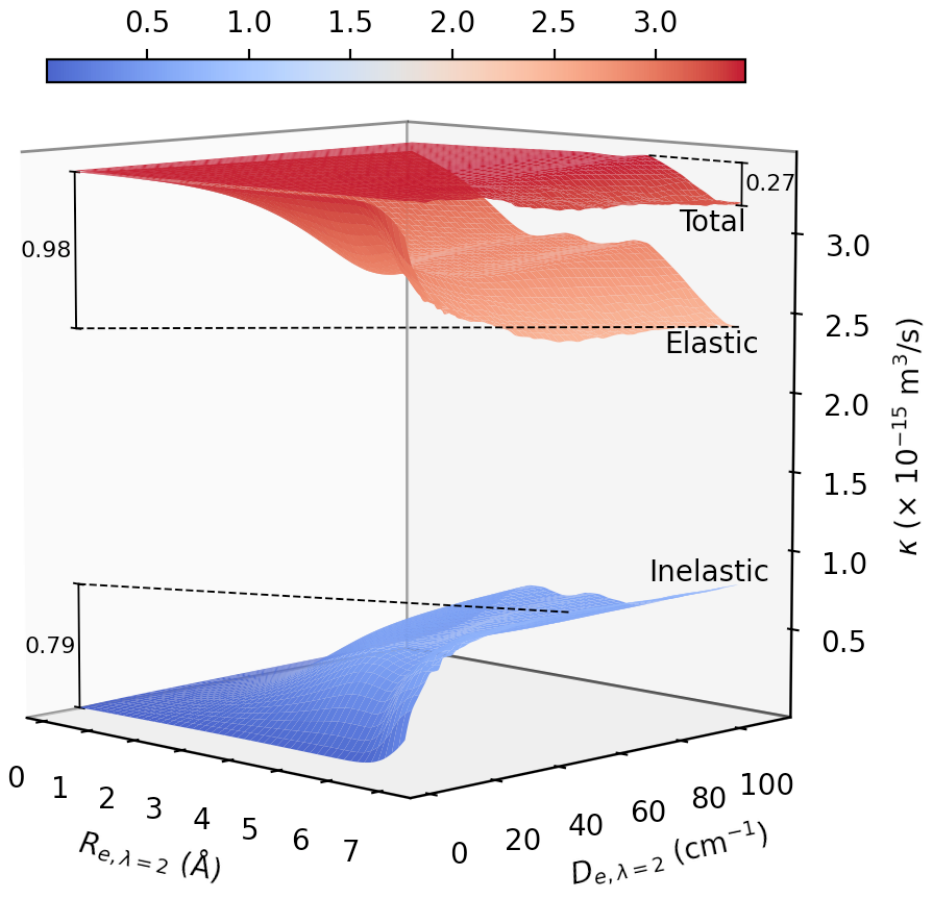}
\caption{Rate coefficients for total, elastic, and inelastic scattering of Rb with N$_2$($j=0$) as functions of $R_{\mathrm{e},\lambda=2}$ and $D_{\mathrm{e},\lambda=2}$ from CC calculations. The inelastic rate coefficients increase up to 30\% of the elastic rate coefficients with the variations in interaction anisotropy. However, the total rate coefficients are largely insensitive to the variations due to the mutual compensation of the inelastic and elastic rate coefficients.}
\label{figure4}
\end{figure}

\begin{figure}[b]
    \centering
    \includegraphics[width=0.48\textwidth]{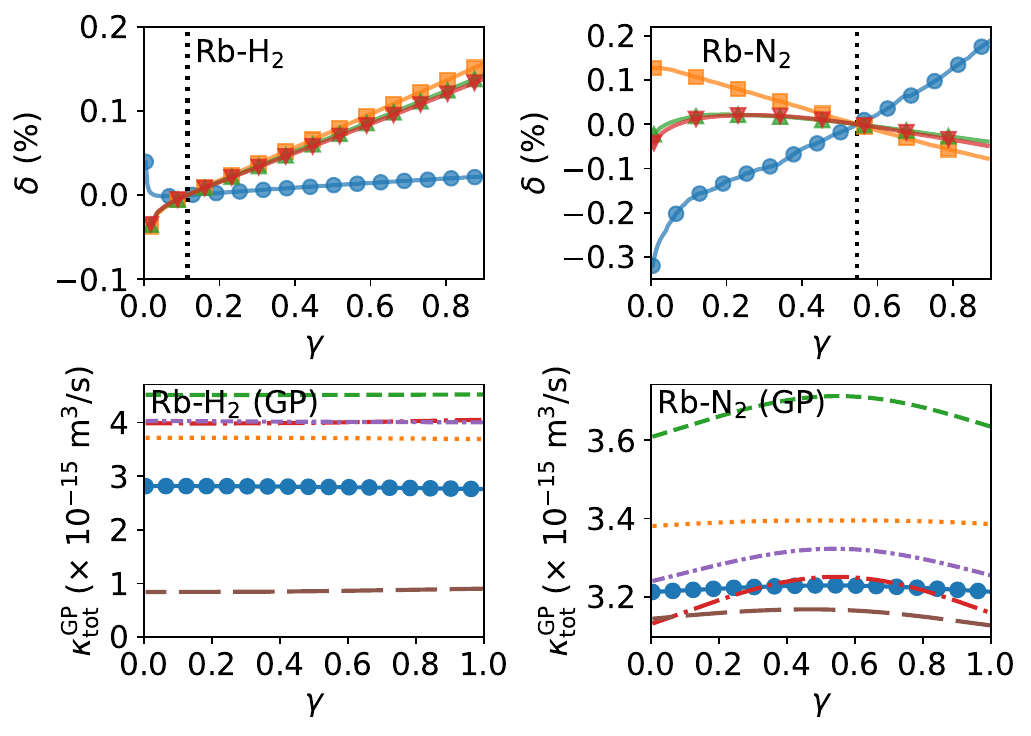}
\caption{Upper panels: The relative deviation~(\ref{eq:rel_err}) as a function of $\gamma= C_{6,\lambda=2}/C_{6,\lambda=0}$ for collisions of Rb with H$_2$ (left) and N$_2$ (right) initially in the rotational state
$j=0$ (circles), $j=2$ (squares), $j=4$ (triangles), $j=6$ (inverted triangles). The reference values of $\gamma$ are shown by the vertical dotted lines.
Lower panels: Rate coefficients $\kappa_{\rm{tot}}$ from the GP models as functions of $\gamma$. 
 The lines with the symbols represent $\kappa_{\rm{tot}}$ marginalized over $R_{\text{e}, \lambda = 2}, D_{\text{e}, \lambda = 2}, D_{\text{e},\lambda=0}$ and $B_{\rm r}$. 
   The dashed and dotted lines without symbols show samples of ${\kappa}_{\rm tot} = f(R_{\text{e}, \lambda = 2}, D_{\text{e}, \lambda = 2}, C_{6,\lambda=2}, D_{\text{e},\lambda=0}, B_{\rm r} )$ for random combinations of 
$R_{\text{e}, \lambda = 2}, D_{\text{e}, \lambda = 2}, D_{\text{e},\lambda=0}$ and $B_{\rm r}$. 
The GP models are trained by the CC calculations performed with the reduced mass, $C_{6,\lambda=0}$ and $R_{{\rm e},\lambda=0}$ corresponding to the \rbh\ (left) and \rbn\ (right) collision systems.
}
\label{figure5}
\end{figure} 

\section{Results}

We begin by calculating rate coefficients for total \rbh$(j)$\ and \rbn$(j)$\ scattering at room temperature using three methods: (i) rigorous multi-channel CC calculations with the best fits of full PES described in Section~\ref{sec:cc}; (ii) single-channel scattering calculations using only the isotropic part $V_{\lambda = 0}(R)$ of the interaction potentials; (iii) semi-classical Jeffreys-Born approximation~(\ref{eq:JBJB}) based on the long-range part of $V_{\lambda = 0}(R\rightarrow \infty)$. The results are presented in Table~\ref{tab:rate} for molecules initially in different rotational states $j=$ 0 to 6. 
 
We note a very good agreement between \colrate{\rm{tot}} from full CC calculations and \colrate{\rm{SC}} based on the semi-classical Jeffreys-Born approximation for \rbn\ collisions but not for \rbh\ collisions. This illustrates that quantum scattering of Rb by N$_2$ is qualitatively different from \rbh\ collisions, as a consequence of the dramatic difference in the corresponding PES and reduced masses. In the present work, we use the PES for \rbn\ and \rbh\ collision complexes as a starting point for our analysis and perform calculations with the reduced mass for \rbn\ and \rbh\ collisions.

We observe a remarkable agreement between \colrate{\rm{tot}} from full CC calculations and \colrate{\rm{single}} from single-channel calculations for both \rbn\ and \rbh\ collisions and for all rotational states of the molecule considered. 
The results for $\kappa_{\rm single}$ are obtained from the cross sections computed by integrating the single-channel differential equation that corresponds to Eq. (\ref{cc}) with 
all $V_{\lambda > 0}(R)$ in Eq. (\ref{legendre}) set to zero. 
Thus, the PES used for the computation of $\kappa_{\rm single}$ ignores entirely the anisotropy of the atom - molecule interaction. 
The agreement between $\kappa_{\rm tot}$ and  $\kappa_{\rm single}$
 is particularly interesting for \rbn\ collisions and for collisions of molecules in excited states, which exhibit significant rates of inelastic scattering. 
In the following sections, we examine in detail the response of thermal rate coefficients to variations in the interaction anisotropy in order to elucidate the extent and generality of the agreement between the results of multi-channel and single-channel quantum scattering calculations of $\kappa_{\rm tot}$.

\begin{table}[!htbp]

    \caption{\label{tab:rsd}
    Relative standard deviations ($\mathcal{R}_{\text{w}}$ and $\mathcal{R}_{10\%}$ in \%) of the distributions of $\kappa_{\rm{tot}}$ for total Rb + M collisions due to variations in the well depth $D_{\text{e}, \lambda}$ and the equilibrium distance $R_{\text{e}, \lambda}$ of the Legendre expansion terms of the atom - molecule PES. The parameters are sampled from     
    a wide range $R_{\text{e}, \lambda} \in [0.25,7]~\text{\AA}$, $D_{\text{e}, \lambda} \in [1,110]~\mathrm{cm}^{-1}$ yielding $\mathcal{R}_{\text{w}}$, and the $\pm$10\% deviation from the reference values given by the MLR fits (\ref{eq:mlr}), yielding 
  $\mathcal{R}_{10\%}$.}
    
    \begin{ruledtabular}
        \begin{tabular}{lllll}
        M & Parameter & Reference value & $\mathcal{R}_{\text{w}}$ (\%) & $\mathcal{R}_{10\%}$ (\%) \\ 
        \hline
        \multirow{4}{*}{\hh} 
            & $D_{\text{e}, \lambda=2}$ & 3.51 cm$^{-1}$ & \multirow{2}{*}{3.4} & \multirow{2}{*}{0.004} \\ 
            & $R_{\text{e}, \lambda=2}$ & 5.52 \AA & & \\ 
            & $D_{\text{e}, \lambda=0}$ & 7.27 cm$^{-1}$ & \multirow{2}{*}{57.9} & \multirow{2}{*}{10.2} \\ 
            & $R_{\text{e}, \lambda=0}$ & 6.39 \AA & & \\ 
        \midrule
        \multirow{4}{*}{\nn} 
            & $D_{\text{e}, \lambda=2}$ & 11.94 cm$^{-1}$ & \multirow{2}{*}{1.4} & \multirow{2}{*}{0.08} \\ 
            & $R_{\text{e}, \lambda=2}$ & 5.84 \AA & & \\ 
            & $D_{\text{e}, \lambda=0}$ & 19.97 cm$^{-1}$ & \multirow{2}{*}{5.9} & \multirow{2}{*}{1.5} \\ 
            & $R_{\text{e}, \lambda=0}$ & 6.32 \AA & & \\ 
        \end{tabular}
    \end{ruledtabular}
\end{table}

It is instructive to compare the present calculations with the previously published results for collision rate coefficients averaged both over the collision energies and the initial states of the molecules in the thermal gas~\cite{shenCrosscalibrationAtomicPressure2023,klosElasticGlancingangleRate2023}. 
The thermally averaged total rate coefficients are defined as
\begin{eqnarray}
\bar{\kappa}(T) = \sum_j p_j G_j \kappa_j(T),
\label{eq:popavg}
\end{eqnarray}
where \(\kappa_j(T)\) is the total rate coefficient from Eq.~(\ref{eq:cal_rate}) for molecules with the initial rotational state $j$, $p_j$ is the Boltzmann factor $(2j+1)\exp{(-E_j/(k_B T))}/Q_{\rm rot}(T)$, \(Q_{\rm rot}(T)=\sum_j (2j+1)G_j\exp(-E_j/(k_B T))\) is the rotational partition function, and $E_j=B_r j(j+1)$ is the rotational energy of a diatomic molecule in rotational quantum state $j$. The nuclear spin degeneracy for the ortho- and para-isomers is $G_j=(2-(-1)^j)$ for \hh\ and $G_j=(9+3(-1)^j)/2$ for \nn.
Table~\ref{tab:popavg} compares the results of the present work with the rate coefficients thus computed for  \rbh\ and  \rbn\ collisions. 
We note that we average over the rates for both even-$j$ and odd-$j$ rotational states of the molecules.
The present calculations of the total rate coefficients for \rbh\ collisions agree 
with the results from Ref.~\cite{shenCrosscalibrationAtomicPressure2023} 
to within 0.06\%. 
The present calculations for \rbn\ agrees with the results from Ref.~\cite{klosElasticGlancingangleRate2023} to within 1.1\%.
Our results illustrate that the total rate coefficients for para-nuclear spin isomers  are not substantially different from those for ortho-nuclear spin isomers. 
The present calculations for \rbh\ collisions differs from the results from Ref.~\cite{klosElasticGlancingangleRate2023} by 8.1\%.
The present work and Ref.~\cite{klosElasticGlancingangleRate2023} used 
different PESs for \rbh. Since \rbh\ interactions are largely isotropic, this difference illustrates 
the sensitivity of the thermally averaged total rate coefficients to variations in the isotropic part of the PES, consistent with the results shown in Table~\ref{tab:rsd}.

\subsection{Invariance to short-range anisotropy}
\label{sec:short}

To analyze the effect of anisotropy, we first compute the dependence of the total collision rate coefficients on $D_{\text{e}, \lambda = 0}$ and $D_{\text{e}, \lambda = 2}$ using direct CC calculations as described in Section~\ref{sec:cc}. 
We vary either $D_{\text{e}, \lambda = 0}$ or $D_{\text{e}, \lambda = 2}$ to compute the deviation of the resulting rate coefficient $\kappa^\prime_{\rm tot}$ from the corresponding value in Table~\ref{tab:rate}.  Figure \ref{figure2} (top panels) illustrates the relative deviation
\begin{equation} \label{eq:rel_err}
    \delta=\frac{\kappa^{\prime}_{\text{tot}}-\kappa_{\mathrm{tot}}}{\kappa_{\mathrm{tot}}} \times 100 \%
\end{equation}
as a function of either $D_{\text{e}, \lambda = 0}$ or $D_{\text{e}, \lambda = 2}$ for \rbh($j=0$)\ and \rbn($j=0$) collisions. 
For these calculations, when $D_{\text{e}, \lambda = 0}$ or $D_{\text{e}, \lambda = 2}$ is varied, all other parameters of PES and the molecular constants are fixed to the values corresponding to the specific collision system under examination. 

The results show that $\kappa_{\rm tot}$ are sensitive to changes in $D_{\text{e}, \lambda = 0}$
but remain nearly constant over a wide range of $D_{\text{e}, \lambda = 2}$. More specifically, 
the relative changes $\delta$ remain within 10\% for \rbh($j=0$)\ collision and 3\% for \rbn($j=0$)\ collision, across the wide range of  $D_{\text{e}, \lambda = 2}$ from 1 to 110~\wvnum. 
It is instructive to examine the response of total scattering to variation of $D_{\text{e}, \lambda = 2}$ for molecules in excited rotational states, which permit both excitation and relaxation inelastic transitions.  We repeat the calculations of the total rate coefficients for molecules initially in the $j = 2, 4$, and $6$ rotational states. The resulting values of $\delta$ are shown 
in Fig.~\ref{figure2} (lower panels) for a wide range of $D_{\text{e}, \lambda = 2}$. For reference, the values of $D_{\text{e}, \lambda = 2}$ corresponding to the best fits of the corresponding PES are shown in Fig.~\ref{figure2}  by the vertical dotted lines.

To illustrate the lack of sensitivity to $D_{\text{e}, \lambda = 2}$ for a wide range of molecular collision parameters, we analyze the distribution of $\kappa_{\rm tot}$ as a function of $D_{\text{e}, \lambda = 2}$ obtained by random sampling from the 5D GP model described in Section~\ref{sec:stat} and by marginalization of the GP model predictions over $R_{\text{e}, \lambda = 2}, C_{6,\lambda=2}, D_{\text{e},\lambda=0}$ and $B_{\rm r}$. 
The upper panels of Fig.~\ref{figure3} illustrate that the distributions exhibit a significant variance with the randomly sampled parameters, but are squeezed in $D_{\text{e}, \lambda = 2}$. 
Specifically, for each combination of parameters, the total rate coefficients are largely insensitive to variations in $D_{\text{e}, \lambda = 2}\in [0,100]$ cm$^{-1}$. 
The lower panels of Fig.~\ref{figure3} illustrate the dependence of $\kappa_{\rm tot}$ on the position of the minimum $R_{\text{e}, \lambda = 2}$. It can be seen that a very big change in $R_{\text{e}, \lambda = 2}$ (corresponding to a very big change in the interaction anisotropy) is required to induce a change of $\kappa_{\rm tot}$ exceeding 1 \% as shown in the shaded regions. 
We note that these distributions reflect the dependence of $\kappa_{\rm tot}$ on $D_{\text{e}, \lambda = 2}$ for a range of molecules with different rotational constants and a distribution of PES with widely different anisotropies as determined by both $R_{\text{e}, \lambda = 2}$ and $D_{\text{e}, \lambda = 2}$.

While $\kappa_{\text{tot}}$ appears to be invariant to changes in short-range anisotropy, the rate coefficients for elastic and inelastic scattering change significantly with the interaction anisotropy. 
This is illustrated in Fig. \ref{figure4} that depicts $\kappa_{\text{el}}$, $\kappa_{\text{inel}}$ and  $\kappa_{\text{tot}}$ for Rb -- N$_2(j=0)$ collisions obtained with direct CC calculations as functions of $R_{\text{e}, \lambda = 2}$ and $D_{\text{e}, \lambda = 2}$.  Figure \ref{figure4} shows that the elastic rate coefficients decrease as the inelastic rate coefficients grow, restricting the variation of $\kappa_{\text{tot}}$ despite the wide range of $R_{\text{e}, \lambda = 2}$ and $D_{\text{e}, \lambda = 2}$ considered. We note that the PES corresponding to the values of $R_{\text{e}, \lambda = 2}$ and $D_{\text{e}, \lambda = 2}$ from the opposite limits of the corresponding ranges are qualitatively different. Thus, PES with small values of $D_{\text{e}, \lambda = 2}$ exhibit one minimum in the T-shaped configuration (defined as $\theta = 90$ degrees), whereas large values of $D_{\text{e}, \lambda = 2}$ introduce an additional minimum at the linear atom - molecule ($\theta = 0$ degrees) geometry.

Table~\ref{tab:rsd} quantifies the variation of $\kappa_{\text{tot}}$ with the change of both the isotropic part and the anisotropy of PES at short range. 
To obtain the results in Table~\ref{tab:rsd}, we vary 
the equilibrium distance $R_{\rm{e}}$ for both the  isotropic $\lambda = 0$ and anisotropic $\lambda = 2$ contributions to the interaction potential. This results in a distribution of  $\kappa_{\text{tot}}$. We compute the distribution of $\kappa_{\text{tot}}$ by sampling from a wide range of PES parameters $R_{\text{e}, \lambda} \in  [0.25, 7]$~\AA\ and 
$D_{\rm{e}, \lambda} \in [1, 110]$~\wvnum\ and the distribution of $\kappa_{\text{tot}}$ by sampling from the range of $R_{\text{e}, \lambda}$ and $D_{\text{e}, \lambda}$ encompassing $\pm 10$ \%
deviations from the values in the best MLR~(\ref{eq:mlr}) fits of the corresponding \textit{ab initio} potentials for \rbh\ and \rbn.  
The first sampling is designed to explore the response to $\kappa_{\text{tot}}$ to extreme changes of PES. 
The second sampling is designed to vary the PES within a range closer to an expected error of quantum chemistry calculations. We note that the $\pm 10$ \% variation of the PES parameters, though considered here as small, significantly exceeds the estimated uncertainties of the {\it ab initio} calculations and the fits of the interaction potentials~\cite{medvedevInitioInteratomicPotentials2018,klosElasticGlancingangleRate2023}.

\subsection{Invariance to long-range anisotropy}
\label{sec:long} 
In this section, we examine the response of the total thermal rate coefficients to the anisotropy of PES at long range, quantified by the parameter
$   \gamma = C_{6,\lambda=2}/C_{6,\lambda=0}$, where $C_{6,\lambda=0}$ and $C_{6,\lambda=2}$ are the leading coefficients in expansion~(\ref{LR}) of $V_{\lambda = 0}$ and $V_{\lambda =2}$, respectively. 
We fix \(C_{6,\lambda=0}\) to the values of the best fit of the corresponding {\it ab initio} potentials and generate a series of PES in the range of $\gamma$ from 0 to 1 by varying $C_{6,\lambda=2}$. 
The upper panels of Fig. \ref{figure5} show the departure of $\kappa_{\rm tot}$ from the corresponding values in Table~\ref{tab:rate} as functions of $\gamma$ for collisions of Rb with H$_2$ and N$_2$ in different rotational states. The results show that the long-range anisotropy, even when $C_{6,\lambda=2} = C_{6,\lambda=0}$ does not alter $\kappa_{\rm tot}$ by more than a small fraction of one percent. 

The lower panels of Fig. 5 show the dependence of $\kappa_{\rm tot}$ on $\gamma$ predicted by the GP model described in Section~\ref{sec:stat}. 
The lines with the symbols show the predictions marginalized over $R_{\text{e}, \lambda = 2}, D_{\text{e},\lambda=2}, D_{\text{e},\lambda=0}$ and $B_{\rm r}$, whereas the dashed and dotted lines without symbols 
show model predictions for random samples from various combinations of $R_{\text{e}, \lambda = 2}, D_{\text{e},\lambda=2}, D_{\text{e},\lambda=0}$ and $B_{\rm r}$. 
It is clear from these results that $\kappa_{\rm tot}$ remains invariant to changes in $\gamma$ for a wide range of PES with different anisotropies and a wide range of rotational constants of the diatomic molecule.

\subsection{Single-channel vs multi-channel scattering calculations}
\label{sec:single_mult}
The numerical complexity of multi-channel scattering calculations scales as $O(N^3)$ where $N$ is the number of coupled channels. Converged scattering calculations of cross sections for the total thermal rate coefficients for \rbn\  collisions at room temperature require 
$N=742$ coupled channels. Based on the results in Sections~\ref{sec:short} and \ref{sec:long}, we argue that single-channel ($N = 1$) quantum scattering calculations using only the isotropic part $V_{\lambda = 0}(R)$ of PES can 
achieve similar accuracy for total thermal rate coefficients as multi-channel CC calculations.

\begin{figure}[b]
    \includegraphics[width=0.48\textwidth]{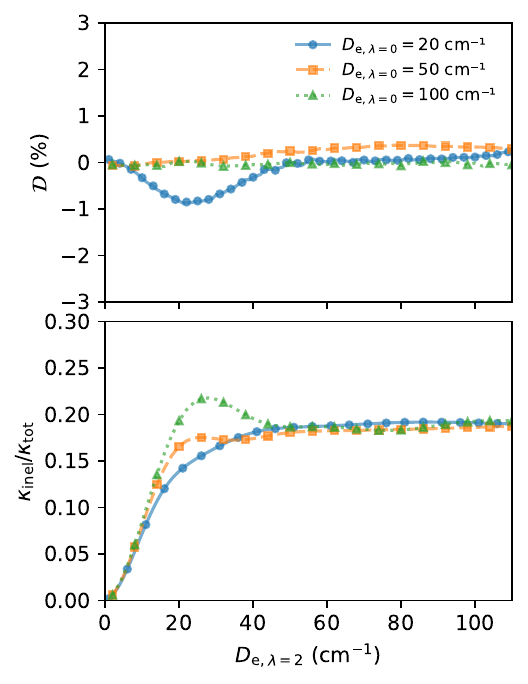}

\caption{
    Relative deviations~(\ref{eq:dev}) and relative contribution of inelastic scattering to the total scattering rate coefficient as functions of $D_{\mathrm{e}, \lambda=2}$ for collisions of Rb with N$_2$($j=0$) for $D_{\mathrm{e}, \lambda=0}=$20 (circles), 50 (squares) and 100~cm$^{-1}$ (triangles).
}
\label{figure6}
\end{figure}

To further support this argument and show that this conclusion extends to a wide range of molecular systems, we express the atom-molecule interaction potentials as
\begin{equation}
    V(R, \theta)=V_0\left(1+\sum_{\lambda=2,4,6} \frac{D_{\text{e}, \lambda}}{D_{\text{e},\lambda=0}} \cdot \frac{\mathcal{F}_\lambda(R)}{\mathcal{F}_0(R)} P_\lambda(\cos \theta)\right),
\end{equation} where $\mathcal{F}_\lambda(R)$ is the radial term in Eq.~(\ref{fterm}).
The term with $\lambda=2$ is scaled by the ratio $D_{\text{e},{\lambda = 2}}/D_{\text{e},{\lambda 0}}$.
We vary $D_{{\rm e}, \lambda = 2}$ from 1 to 110~\wvnum\, and set $D_{{\rm e}, \lambda = 0}$ to $20, 50, 100$~\wvnum. 
We then perform both converged multi-channel calculations with full PES and single-channel calculations using only $V_{\lambda = 0}$ 
to compute the total scattering rate coefficients. The relative difference of the single- and multi-channel results is quantified by $\mathcal{D}$, defined as
\begin{equation} \label{eq:dev}
    \mathcal{D} = \frac{\kappa_{\text{single}} - \kappa_{\text{CC}}}{\kappa_{\text{CC}}} \times 100\%,
\end{equation} where $\kappa_{\text{single}}$ and $\kappa_{\text{CC}}$ are the total rate coefficients from the single- and multi-channel CC scattering calculations, respectively. The relative difference is presented in the upper panel of Fig.~\ref{figure6}. 
The results show that the largest differences for $D_{{\rm e}, \lambda = 0}=20, 50, 100$~\wvnum\ are -0.88\%, 0.37\%, and -0.089\%, despite the inelastic rate contributions reaching up to 19\%, 18\%, and 22\% of the total rate coefficients as shown in the lower panel of Fig.~\ref{figure6}. The results suggest that single-channel calculations can yield total rate coefficients with the same accuracy as multi-channel calculations for atom-molecule collisions with strong interaction anisotropy. 

\begin{figure}[b]
    \includegraphics[width=0.48\textwidth]{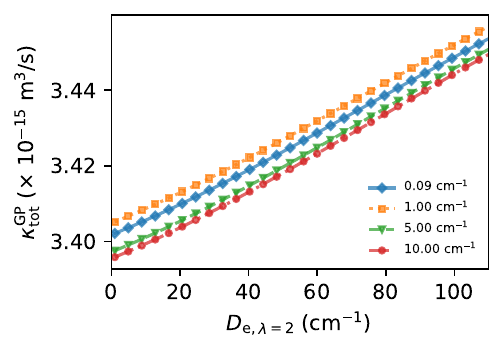}
\caption{Rate coefficients $\kappa_{\text{tot}}$ from the 5D GP model as functions of $D_{\mathrm{e},\lambda=2}$, for the collision systems with rotational constant $B_r =$ 0.09 (blue diamonds), 1.0 (orange squares), 5 (green triangles), and 10~cm$^{-1}$ (red circles). The CC calculations for training and testing the GP model are performed with the reduced mass, $C_{6,\lambda =0}$ and $R_{e,\lambda =0}$ corresponding to the Rb-N$_2$ collision system. Lines with symbols represent $\kappa_{\text{tot}}$ marginalized over all variables except $B_r$ and $D_{\mathrm{e},\lambda=2}$. }
\label{figure7}
\end{figure}

We further illustrate the agreement between single- and multi-channel calculations for molecules 
with different rotational constants $B_{\rm r}$. 
We consider a range of the rotational constant from 0 to 10~\wvnum. 
For reference, the magnitude of the rotational constant for N$_2$  is 1.998~\wvnum~\cite{Bendtsen1974,Lofthus1977}. 
We use the GP model of the rate coefficients for  \rbn($j=0$) collisions to examine the dependence of $\kappa_{\rm tot}$ on \deaniso{}.
Fig.~\ref{figure7} shows the model predictions marginalized over the remaining three variables ($R_{\text{e}, \lambda = 2}$, $C_{6, \lambda = 2}$ and $D_{\text{e}, \lambda = 0}$). 
This accounts for a wide range of PES corresponding to different values of \deiso{} and long-range interaction parameters. 
The results show that the change of the magnitude of $\kappa_{\rm tot}$ remains under 1 \% for all rotational constants considered despite the variation of \deaniso{} over a wide range from 0 to 100~\wvnum. We also observe a very weak dependence of $\kappa_{\rm tot}$ on the rotational constant of the molecule.

\begin{figure*}
    \includegraphics[width=\textwidth]{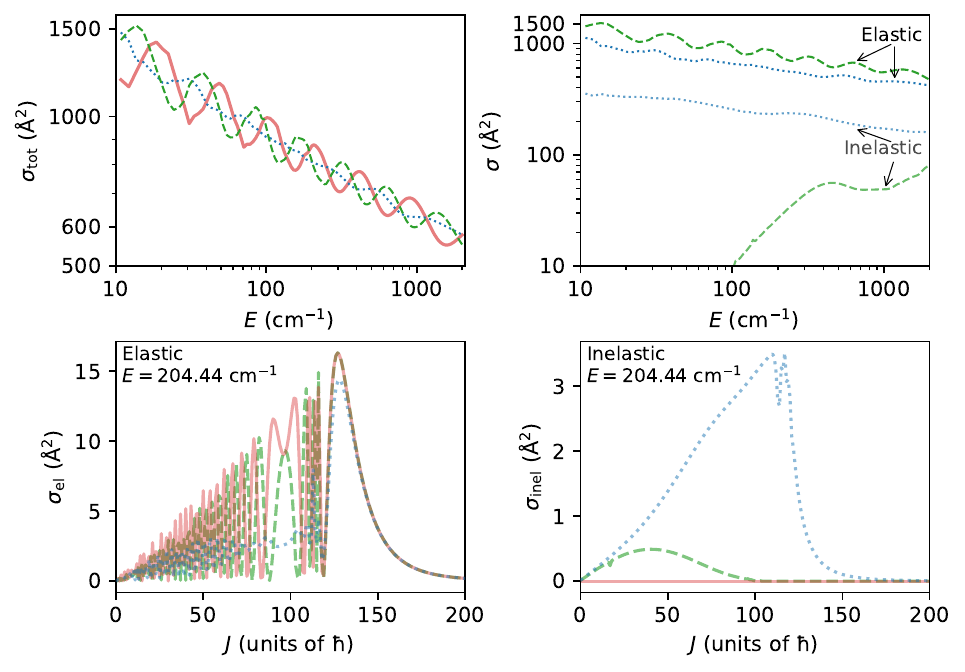}
\caption{Cross sections from CC calculations performed with the reduced mass, $C_{6,\lambda =0}$ and $R_{e,\lambda =0}$ corresponding to the Rb-N$_2$ collision system with the rotational constant $B_r$ varied to 0.09 (blue dots) and 10~cm$^{-1}$ (green dashes), and from single-channel calculations (red solid curves). The $D_{\mathrm{e},\lambda=0}$ and $D_{\mathrm{e},\lambda=2}$ are varied to 100~cm$^{-1}$. Upper panels: Total (left), elastic (right), and inelastic (right) cross sections as functions of collision energy $E$. Lower panels: Elastic (left) and inelastic (right) cross sections as functions of total angular momentum $J$ with the collision energy $E = 204.44$~cm$^{-1}$.}
\label{figure8} 
\end{figure*}

We conjecture that the agreement between the total rate coefficients from single-channel scattering calculations with $V_{\lambda = 0}$ and multi-channel CC calculations is a consequence of the
Maxwell-Boltzmann averaging (\ref{eq:cal_rate}) that diminishes the effect of the glory oscillations and reduces the sensitivity of quantum diffractive scattering to PES at short range so long as the collision velocities do not exceed a certain threshold~\cite{bernsteinExtremaVelocityDependence1962,bernsteinSemiclassicalAnalysisExtrema1963a,childMolecularCollisionTheory2014,guo2024boundariesuniversalitythermalcollisions,boothRevisingUniversalityHypothesis2024}.
To illustrate the effect of thermal averaging, we compute the cross sections as functions of the collision energy for scattering of Rb by molecules with different rotational constants. The results of the single-channel and multi-channel CC calculations are shown in the upper panel of Fig.~\ref{figure8}. 
The energy dependences of the cross sections from the single-channel calculation and the results of the CC calculations for different rotational constants 
oscillate about the same background value. Thermal averaging of the three different energy dependences yields the rate coefficients within 0.16\%.

This remarkable agreement of the thermally averaged rate coefficients occurs despite drastically different dynamics of atom - molecule collisions for molecules with different rotational constants, as illustrated in Fig. \ref{figure8}. The upper right panel of Fig. \ref{figure8} shows the energy dependences of the elastic and inelastic cross sections for molecules with $B_{\rm r}=0.09$ and $B_{\rm r}=10$~\wvnum.     
We observe that the amplitude of the glory oscillations decreases with decreasing $B_{\rm r}$, as a result of the enhancement of inelastic scattering. 
The lower panels of Fig.~\ref{figure8} further show the elastic and inelastic scattering cross sections at a fixed collision energy $E=k_BT=204.44$~\wvnum\ as functions of the total angular momentum $J$. We observe that, for low values of $J$, the elastic scattering cross sections decrease and the inelastic scattering cross sections increase with decreasing $B_{\rm r}$. The total scattering cross sections for large $J$ are universal for collisions of molecules with different rotational constants, consistent with the observation in Ref.~\cite{boothUniversalityQuantumDiffractive2019}. 
The glory oscillations arise from stationary-phase interference between different partial wave states and can be influenced by near-threshold bound or quasi-bound states~\cite{Massey1933,massey1933free,massey1934free,bernsteinExtremaVelocityDependence1962,bernsteinSemiclassicalAnalysisExtrema1963a,childMolecularCollisionTheory2014,boothRevisingUniversalityHypothesis2024}. The glory oscillations are suppressed by inelastic scattering in the presence of open collision channels, leading to the disappearance of the glory oscillations from the energy dependence of cross sections shown in Figure~\ref{figure8}.
At the same time, the results in Fig.~\ref{figure8} show that the increase in inelastic cross sections with decreasing $B_{\rm r}$ compensates for the decrease in elastic cross sections, consistent with the observation in Fig.~\ref{figure4}.

\begin{figure*}

    \includegraphics[width=\textwidth]{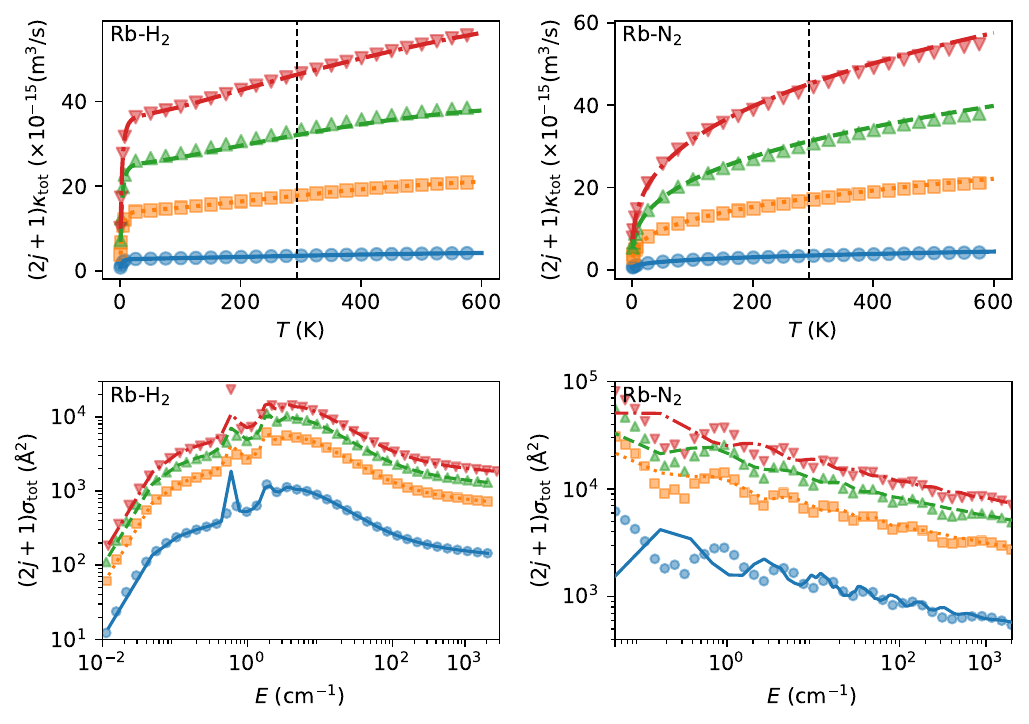}
\caption{  Temperature (top panels) and collision energy (lower panels) dependence of $(2j+1)\kappa_{\text{tot}}$ and $(2j+1)\sigma_{\text{tot}}$ for the collisions of Rb with H$_2$ (left panels) and N$_2$ (right panels) initially in the rotational state $j=0$ (circles), $j=2$ (squares), $j=4$ (triangles), $j=6$ (inverted triangles), from rigorous CC calculations using the full anisotropic potentials (curves) and only the isotropic $\lambda = 0$ contribution to the PES (symbols). The dashed vertical line indicates $T=294.15$~K. }

\label{figure9} 
\end{figure*}

To illustrate the temperature dependence of the invariance of the collision rate coefficients to the interaction anisotropy,
we compute the total rate coefficients for \rbh\ and \rbn\ collisions for the range of temperatures from 1~K to 600~K, as shown in the top panels of Fig.~\ref{figure9}. 
The results show that the total rate coefficients calculated using the full anisotropic potentials agree with those calculated using the isotropic potentials over the temperature range considered. 
The small difference shows that the contribution of inelastic rate coefficients to total rate coefficients is within 1\% for Rb -- H$_2(j=0,2,4,6)$ and Rb -- N$_2(j=0,2,4,6)$ collisions at 294.15~K. 
We note that the difference increases with $T$, up to 4\% for Rb -- N$_2(j=6)$ collisions at 600~K. 
Similar trends for Li-N$_2$ collisions are reported in Ref.~\cite{klosElasticGlancingangleRate2023}. 
We observe good agreement between the total cross sections for \rbh\ and \rbn\ computed using the full anisotropic potentials and the isotropic potentials except for the resonances region of \rbn\ scattering where the anisotropy produces noticeable deviations, as shown in the lower panels of Fig.~\ref{figure9}. 
We note that the Maxwell-Boltzmann distribution in Eq.~(\ref{eq:cal_rate}) peaks near $E_p=k_BT$, with mean $2k_BT$ and width $\sqrt{2}\,k_BT$. 
For H$_2$, $B_{\rm r}=60.85$~cm$^{-1}$~\cite{STAVROS200833} and the first inelastic threshold lies near $E_{j\to j+2}=365.10$ cm$^{-1}$. The peak of the Maxwell-Boltzmann probability denisty is $E_p=k_BT=347.50$ cm$^{-1}$ at 500 K. The difference between $E_{j \to j+2}$ and $E_p$ explains the rigid behaviour of \rbh\ over 1 K to 500 K. For \rbn, $B_r=1.998$~cm$^{-1}$~\cite{huber2013molecular} and $E_{0\to2}=12$ cm$^{-1}=17.28$ K. The \rbn\ collisions above 18 K can access the open inelastic channel $j \rightarrow j+2$.  
The total rate coefficients become smooth functions of $T$ once averaged over the collision energy. 
Anisotropy-induced effects on total cross sections close to $E_{j \rightarrow j+2}$ are strongly surpressed unless $k_B T$ lies deep in the threshold regime. 
For $T > 40 \mathrm{~K}$, the dominant energies already exceed the energy of the ${j \rightarrow j+2}$ excitation for $\mathrm{N}_2$ by multiple widths. The total rate coefficients are thus  dominated by the smooth, long-range part of the interaction. Any low-energy resonances or partial-wave onsets change $\sigma_j(E)$ locally but do not survive the thermal averaging in $\kappa_j(T)$~(\ref{eq:cal_rate}), leaving only minor deviations at the lowest temperatures.

\section{Summary}
\label{sec:sum}
We have presented a comprehensive analysis of the response of thermally averaged total rate coefficients for atom-molecule collisions to variations in both short-range and long-range anisotropy of the atom - molecule PES. 
We use rigorous quantum scattering theory, the interaction potentials based on high-level electronic structure calculations for \rbn\ and \rbh\ collisions, and the statistical analysis approach based on Gaussian process regression to examine in detail the extent and the origin of the invariance of total collision rate coefficients to changes in the interaction anisotropy.

It is illustrated that the invariance of the total scattering rate coefficients to the interaction anisotropy is independent of the initial rotational state and the rotational constant of the diatomic molecule. To explore the limits of the invariance with respect to the strength of the interaction anisotropy, we examine the response of the total rate coefficients to the wide variations of PES. 
Our calculations illustrate that the total scattering rate coefficients change by less than 1 \% as \deaniso{} is increased by a factor of 10 and the long-range anisotropy parameter $\gamma=C_{6,\lambda = 0}/C_{6,\lambda = 2}$ is increased by a factor of 20 from the values corresponding to the best fits of PES for \rbn\ and \rbh\ collisions. 
We note that these ranges are much wider than expected uncertainties from electronic structure calculations. For example, the relative uncertainties of well depths of interaction potentials range from 1\% to 15\% in Refs.~\cite{klosElasticGlancingangleRate2023,medvedevInitioInteratomicPotentials2018}.

The invariance of total scattering to the interaction anisotropy suggests that thermal collision rates for total atom-molecule scattering can be computed using only the isotropic part of PES, which requires the integration of a single (uncoupled) differential equation instead of a large number of coupled equations for scattering calculations with full PES. 
Our analysis shows that single-channel scattering calculations agree with full CC calculations of total scattering rate coefficients for \rbn$(j=0,2,4,6)$\ and \rbh$(j=0,2,4,6)$\ collisions to within 0.92\%. This agreement remains within 0.88\% for molecules with a wide range of rotational constants, 
despite the inelastic rate contributions reaching up to 29\% of the total collision rate coefficients.

Our scattering calculations systematically illustrate that the invariance of total scattering rate coefficients is due to the mutual compensation of elastic and inelastic scattering rate coefficients. 
The agreement between the results of single- and multi-channel scattering calculations is shown to be due to the Maxwell-Boltzmann averaging~(\ref{eq:cal_rate}). Thermal averaging also reduces the dependence of the total scattering rate coefficients on the rotational constant of the molecule. We have shown that the glory oscillations in the scattering cross sections for molecules with small rotational constants are significantly suppressed by the enhancement of inelastic scattering. While the glory oscillations remain significant for molecules with large rotational constants, thermal averaging reduces the influence of the glory oscillations, consistent with the observations in Refs.~\cite{bernsteinExtremaVelocityDependence1962,bernsteinSemiclassicalAnalysisExtrema1963a,childMolecularCollisionTheory2014,guo2024boundariesuniversalitythermalcollisions,boothRevisingUniversalityHypothesis2024}.  
We note that the numerical examples in this work are restricted to atom -- homonuclear diatom systems with $-C_6/R^6 - C_8/R^8 - C_{10}/R^{10}$ long-range interaction and even ($\lambda=2,4,6$) Legendre expansion terms of the anisotropic PES. Extensions to ion--neutral, molecule--molecule, and strongly polar/heteronuclear systems which may feature $R^{-4}$ or $R^{-3}$ leading terms and significant odd-$\lambda$ contributions to the interaction anisotropy are left for future study.

\section*{Acknowledgments}
We acknowledge the financial support from the Natural Sciences and Engineering Research Council of Canada (NSERC grants RGPIN-2019-04200, RGPAS-2019-00055) and the Canadian Foundation for Innovation (CFI project 35724). 
This work was conducted at the Center for Research on Ultra-Cold Systems (CRUCS) and was supported in part by computational resources and services provided by Advanced Research Computing at the University of British Columbia. 
X.G., K.W.M., and R.V.K. acknowledge support from the Deutsche Forschungsgemeinschaft funded GRK 2717 grant for the Research Training Group "Dynamics of Controlled Atomic and Molecular Systems (DynCAM)".  

\newpage
\bibliography{ref}%

\end{document}